\documentclass[%
 reprint,
 amsmath,amssymb,
 aps,
]{revtex4-2}

\usepackage{graphicx}
\usepackage{dcolumn}
\usepackage{bm}

\usepackage[T1]{fontenc} 
\usepackage{booktabs} 
\usepackage{ dsfont }
\usepackage{float}

\usepackage{calrsfs}
\DeclareMathAlphabet{\pazocal}{OMS}{zplm}{m}{n}
\usepackage{xcolor}
\usepackage{pict2e}
\usepackage[percent]{overpic}
\usepackage{ upgreek }
\usepackage{xfrac}
\usepackage{abraces}
\usepackage{braket}
\usepackage{stmaryrd}
\usepackage{amsmath,empheq}
\usepackage{nicefrac}

\begin{document} 

\preprint{APS/123-QED}

\title{DALI sensitivity to streaming axion dark matter}
\author{Javier De Miguel$^{1,2,3}$}
 \email{javier.miguelhernandez@riken.jp}

\author{Abaz Kryemadhi$^{4}$}%


\author{Konstantin Zioutas$^{5}$}%


  \affiliation{
$^{1}$The Institute of Physical and Chemical Research (RIKEN), \\
Center for Advanced Photonics, 519-1399 Aramaki-Aoba, Aoba-ku, Sendai, Miyagi 980-0845, Japan}
\affiliation{$^{2}$Instituto de Astrof\'isica de Canarias, E-38200 La Laguna, Tenerife, Spain}
\affiliation{$^{3}$Departamento de Astrof\'isica, Universidad de La Laguna, E-38206 La Laguna, Tenerife, Spain}

\affiliation{$^{4}$Dept. Computing,Math \& Physics, Messiah University, Mechanicsburg PA 17055, USA}

\affiliation{$^{5}$Physics Department, University of Patras, GR 26504, Patras-Rio, Greece}

\vspace{5pt}
\author{On behalf of the DALI Collaboration}
\date{\today}

\begin{abstract}
Dark matter substructures emerge naturally in a scenario in which the axion angular field acquires propagating degrees of freedom in a post-inflationary Universe. The DALI experiment is a new generation wavy dark matters interferometer currently in prototyping. Although DALI's main objective is to explore the virialized dark matter in our Galaxy, to a large degree in this article we explore the prospect for detection of fine grained streams made of axions that would enter the Solar System, as suggested by previous work. We find that DALI will have a sensitivity to encounters with these coherent objects that fully spans the window defined by  the coupling strength to photons of representative axion models over a stacked bandwidth of two decades in mass.
\end{abstract}

\maketitle



\section{Introduction}
According to the standard cosmological model, more than 4/5 of the matter in the Universe is non-luminous. Although we have strong evidence of its existence, dark matter (DM) has not been directly detected \cite{ 1933AcHPh...6..110Z, 1970ApJ...159..379R, 2018RvMP...90d5002B}. The axion is a boson emerged in the quantum chromodynamics (QCD) solution to the problem of conservation of charge and parity symmetry in the strong interaction \cite{PhysRevLett.38.1440, PhysRevLett.40.223, PhysRevLett.40.279}. Within a cosmology-compatible parameter space, the axion is a compelling DM candidate \cite{ABBOTT1983133, DINE1983137, PRESKILL1983127}. Although numerous efforts to explore this sector have been made in the past decades, including Refs. \cite{PhysRevD.105.035022, PhysRevD.98.103015, Marsh_2017, PhysRevLett.118.011103, 2014PhRvL.113s1302A, Straniero:2015nvc, 2022JCAP...10..096D, 2022JCAP...02..035D, REGIS2021136075, 2021Natur.590..238B, EHRET2010149, PhysRevD.88.075014, EJLLI20201,  PhysRevD.92.092002,PhysRevLett.59.839, PhysRevD.42.1297, PhysRevLett.104.041301,PhysRevD.97.092001,PhysRevLett.128.241805,doi:10.1063/5.0098783,2021Natur.590..238B, CAST:2017uph, CAST:2020rlf, Foster:2020pgt, Darling:2020uyo, Darling:2020plz, PhysRevLett.121.261302, PhysRevD.99.101101, PhysRevD.103.102004, MCALLISTER201767, doi:10.1126/sciadv.abq3765, AxionLimits, Quiskamp:2023ehr}, no direct detection of the pseudoscalar has been made.

The standard halo model assumes a smooth distribution of DM particles with a Maxwellian velocity (about 300 km s$^{-1}$). The cold and collisionless nature of DM also suggest fine grained streams with small primordial velocity dispersion, about $10^{-17}c$ for axions~\cite{Sikivie:1995, SikivieVelo:1999, Vogelsberger:2011}.
References~\cite{Vogelsberger:2011, Stucker:2020, Vogelsberger:2020} illustrate that up to $10^{14}$ streams with small velocity dispersions might exist in our Galaxy, where the DM density at any given point arises from the collective densities of these streams; while the number of "massive" streams would be about $10^6$. It is important to note that these streams are distinct from tidal streams, which arise from the disruption of subhalos. 
The small velocity dispersion offers experimental advantages, as sensitivity increases with smaller dispersion. Consequently, these streams may become more accessible to experiments. If detected, they provide a pathway for analyzing the DM within the Galaxy.

Moreover, besides their direct probing appeal in experiments, the small dispersion of individual streams leads to significant gravitational focusing (GF) by solar system bodies, and in particular the self-focusing by the Earth. Detailed discussions on the expected flux amplification resulting from GF can be found in Refs.~\cite{Prezeau:2015, sofue:2020, prdgf:2023}. In our subsequent analysis, we will utilize the findings from Refs.~\cite{prdgf:2023, echostreams:2022} to eventually explore experimental sensitivity.

\begin{figure}[H]
   \centering
 \vspace{2pt}\includegraphics [width=.42\textwidth]{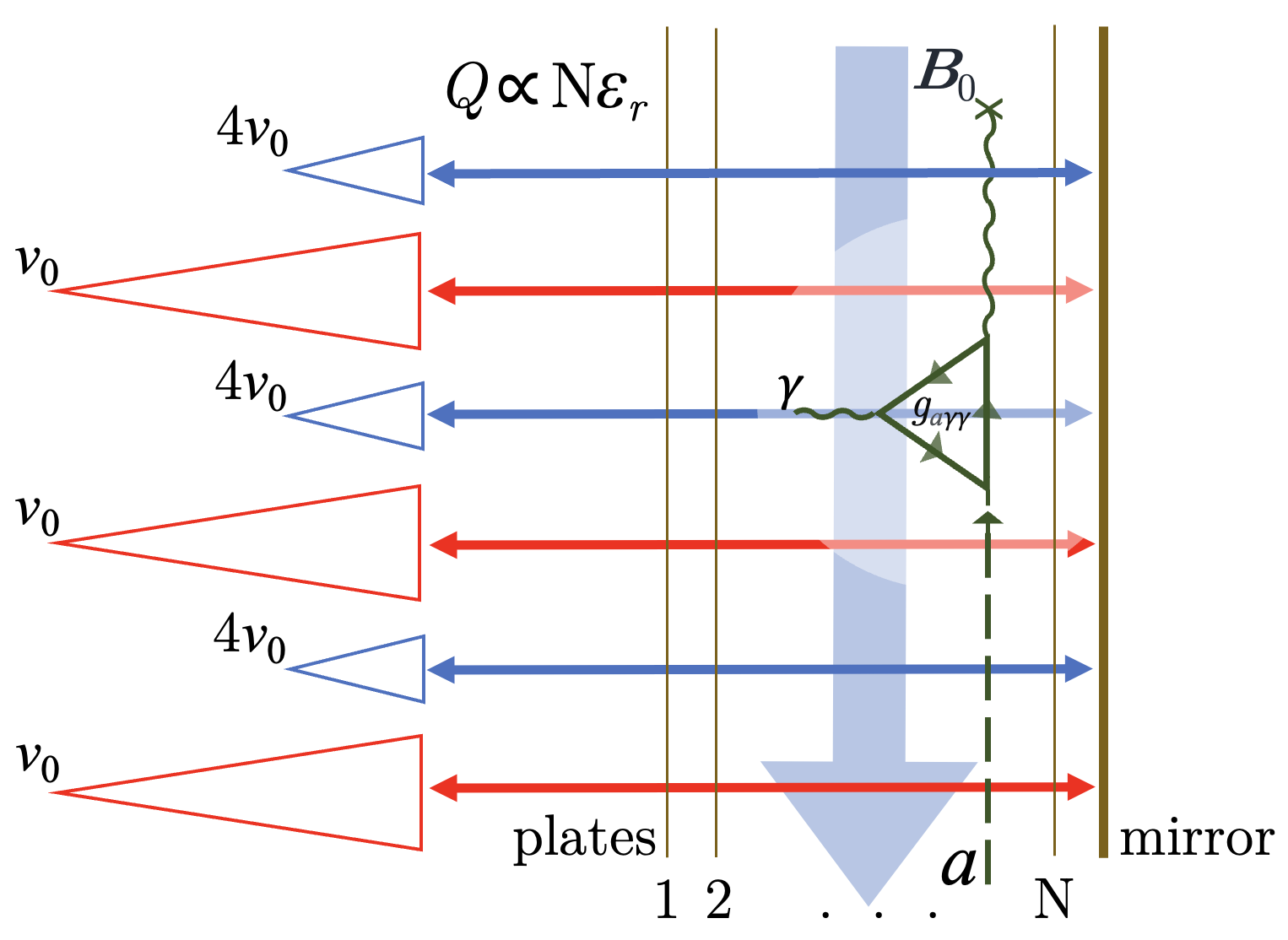}
\caption{A schematic view of the DALI experiment. Ambient axions convert into photons via the inverse Primakoff effect \cite{Primakoff:1951iae}. The system is equipped with a Fabry-Pérot interferometer \cite{1899ApJ.....9...87P} in a magnetic field ($B_0$) to enhance the output power. Microwaves are emitted perpendicularly to the dielectric surfaces regardless of the axion incidence angle. Two resonant frequencies ($\nu_0$) separated by about 4 wavelengths are simultaneously observed with a bandwidth of the order of $\sim$100 MHz each. The quality factor ($Q$) scales with the number of layers (N) and the dielectric permittivity ($\varepsilon_r$).}
\label{fig_0}
\end{figure}

The Dark-photons \& Axion-Like particles Interferometer (DALI) is a new proposal to the search for wavy DM \cite{De_Miguel_2021, DeMiguel:2023nmz, Cabrera:2023qkt, 2024JInst..19P1022H}. DALI is intended to find the axion in a parametric space that is cosmologically tantalizing in a postinflationary picture. In this scenario, where axion emerges after cosmic inflation, the formation of DM substructures is favored,  which particularly motivates the study carried out through this work—e.g., see \cite{MARSH20161}. Crucially, this sector remains poorly explored due to the challenge of probing the axion in high-frequency observations. In addition to a high sensitivity to this range of heavier axions, the DALI project presents some additional advantages, such as enhanced directionality, or a capacity to probe two different resonance frequencies simultaneously, thereby doubling its scan speed. The DALI project is planned in two phases. The second phase consists of an upgrade with a larger and more powerful superconducting solenoid magnet also transferred from the medical resonance imaging industry. A schema is shown in Fig. \ref{fig_0}. DALI is currently in the design and prototyping phase. In this article, we explore the sensitivity of a DALI-like apparatus for the detection of fine grained streams made of axion that could navigate the Solar System according to previous work. The rest of the manuscript is structured as follows. In Sec. \ref{II} we introduce the concept of direct and gravitationally focused fine grained stream. In Sec. \ref{III} we estimate the DALI sensitivity to these events. Conclusions are summarized in Sec. \ref{IV}.

\section{Streaming axion dark matter}\label{II}
Cosmological simulations suggest that the Galaxy may have as many as $10^{14}$ DM streams~\cite{Vogelsberger:2011}. 
As these streams undergo GF in the solar system particularly self-focusing by the Earth, regions of flux enhancements are produced downstream as derived in Refs.~\cite{Prezeau:2015, sofue:2020, prdgf:2023}. In this study we estimate DALI's sensitivity to axion streams for two approaches: first, by analyzing cosmological streams directly before GF onset, and second, by investigating the high density regions formed by the GF.  

In the first scenario, DALI intercepts the DM axion stream just before it passes through the Earth, provided no other solar system bodies intervene, and the incident stream is aligned towards the Earth. Despite the potentially small fraction density of an individual stream, the combination of the small velocity dispersion and the large number of accessible streams compensate for this limitation~\cite{Vogelsberger:2011,prdgf:2023}. Axions from incident streams prior to any gravitational effects would have an inherent dispersion of about $10^{-17}c$~\cite{Sikivie:1995, SikivieVelo:1999, Vogelsberger:2011}. 

In the second scenario, DALI intercepts the DM axion stream just after it passes through the Earth. It's worth noting that only in this configuration does the axion detector becomes exposed to an enhanced flux, potentially reaching up to a factor of about $10^8$ times, due to Earth's gravitational self-focusing (GsF) effect~\cite{sofue:2020, Prezeau:2015, prdgf:2023}. However, to access this high flux with DALI, the stream's laboratory velocity has to be around 20 km s$^{-1}$~\cite{prdgf:2023}. This requirement inevitably leads to a reduction in the number of streams meeting the criteria.  In this scenario, as the stream particles are deflected by the Earth's gravity, the velocity dispersion is increased to about $\sim 10^{-10}c$~\cite{echostreams:2022}.

To facilitate a discussion of sensitivity for both scenarios, the stream information derived from Refs.~\cite{Vogelsberger:2011, Prezeau:2015, sofue:2020, prdgf:2023} is compiled in Table~\ref{tab:stream_table}. This includes estimates of stream density, stream count, density including GsF flux enhancement, and the probabilities for terrestrial experiments to encounter the GsF flux enhancement region.  

For example, according to Table~\ref{tab:stream_table} and Ref.~\cite{Vogelsberger:2011}, a stream with a density of roughly $0.01\,\rho_{0}$—with $\rho_0\sim0.45\, \mathrm{GeVcm^{-3}}$ being the nominal DM density on the Solar System position—is estimated to have a 20\% likelihood of existing within our solar system. Given the substantial density and likelihood of existence such a stream would be a good candidate for direct probing with DALI before GsF. Even though GsF amplifies the density to about $10^6\rho_{0}$ for such a stream, the likelihood of DALI encountering this dense region during a day is only about $2.4\times10^{-10}$.  Direct investigation stands out as the more promising option of search in this case.  
On the other hand, as indicated in Table~\ref{tab:stream_table}, with decreased initial stream density and increased stream count, the probability of encountering a GsF region increases. For instance, the probability that during one day DALI would encounter a GsF region with density of $10\rho_{0}$ becomes about $6\times10^{-3}$. Consequently, investigating the GsF region becomes more promising in this scenario than direct stream investigation. 

\begin{table}
	\centering
	\caption{Stream information before and after gravitational self-focusing (GsF) by the inner Earth.}
	\label{tab:stream_table}
    \begin{ruledtabular}
	\begin{tabular}{lccr} 
		Stream & Stream & GsF density\footnote{The expected density for the specific stream measured in units of $\rho_{0}$, resulting from flux amplification caused by the GsF.} & $P_{\text{day}}$\footnote{The probability that during one day an experiment on Earth will intersect this specific gravitationally focused region.}\\
        density ($\rho_{0}$) & count &  ($\rho_{0}$) &  (GsF encounter)  \\
		\colrule
		$10^{-2}$ & 1 & $10^6$ & $2.4\times10^{-10}$  \\
		10$^{-3}$ & 10 & $10^5$ & $1.2\times10^{-8}$  \\
		10$^{-4}$ & 500 & $10^4$ & $6.0\times10^{-7}$  \\
		10$^{-5}$ & $2\times10^4$ & $10^3$ & $2.4\times10^{-5}$  \\
        10$^{-6}$ & $4\times10^5$  & $10^2$ & $5.0\times10^{-4}$  \\
        10$^{-7}$ & $2\times10^6$  & $10$ & $6.0\times10^{-3}$  \\
         \end{tabular} 
         \end{ruledtabular}
\end{table}

\section{Direct observation of axion streams}\label{III}
Here we calculate the sensitivity of a haloscope of the DALI type to specific axion DM substructures. The axion-like wave is $a(t,x) = a_0\,e^{i(\mathbf{p}\times\mathbf{x}-\omega t)}$; where $\mathbf{p}$ is momentum and  $\mathbf{x}$ any other coordinate, with the axion-like field being its real part. In the low-velocity limit, the axion-induced electric field originating from a local distribution of axions has an amplitude \cite{Millar:2016cjp} 
\begin{equation}
\label{equation_1}
E_0=g_{a\gamma\gamma} B_0 a_0 \simeq 1.3 \times 10^{-12} \frac{\mathrm{V}}{\mathrm{m}} \frac{B_0}{10\, \mathrm{T}} C_{a\gamma\gamma} f^{1/2}_{\mathrm{DM}} \, ,
\end{equation}
where $f_a(\mathrm{x})=f_{\mathrm{DM}}\rho_0$ is the distribution function of axion DM across the DALI vessel, with $f_{\mathrm{DM}}$ a factor introduced to give account for alternative distributions of DM beyond an standard isotropic Halo with density at the Earth's location of $\rho_0\sim0.45\, \mathrm{GeVcm^{-3}}$; while $B_0$ is magnetic field strength and $C_{a\gamma\gamma}$ is the axion-to-photon coupling constant.

The energy flux density across the apparatus is $\pazocal{S} = (\mathrm{\mathbf{E}}_a + \mathrm{\mathbf{E}_{\gamma}} \times \mathrm{\mathbf{H}_{\gamma}})$, with $\mathrm{\mathbf{E}}$ and $\mathrm{\mathbf{H}}$ as the electric and magnetic field for axion and photon. The cycle-averaged power per unit surface reads
\begin{equation}
\label{equation_2}
\frac{\mathrm{P}}{A}=\frac{1}{2}[\mathrm{Re}(\mathrm{\mathbf{E}_{\gamma}} \times \mathrm{Re} (\mathrm{\mathbf{H}_{\gamma}})]=\frac{E^2_0}{2} \, ,
\end{equation}
which, from considering $\mathrm{P}=E_0^2$/($\varepsilon_0 c)$ in Eq. \ref{equation_1} under the assumption that the vessel is in vacuum, results in 
\begin{equation}
\label{equation_3}
\mathrm{P}\simeq 2.2\times10^{-27} \mathrm{W}\frac{A }{\mathrm{m}^2} \left( \frac{B_0}{10\, \mathrm{T}} \right)^2 C^2_{a\gamma\gamma} f_{\mathrm{DM}} \, .
\end{equation}
Introducing in Eq. \ref{equation_3} the Dicke equation \cite{1946RScI...17..268D} for a stabilized radiometer $\mathrm{SNR}=\mathrm{P}/(k_\mathrm{B}T)\sqrt{t/ \mathrm{\delta{\nu}}}$; $g_{a\gamma\gamma}=-2.04(3)\times10^{-16}\,\mathrm{GeV}^{-1}C_{a\gamma\gamma}m_a/\mu \mathrm{eV}$ and the quality factor, $Q$, a dimensionless term of signal power enhancement around a resonant frequency, the  instrument sensitivity becomes
\begin{equation}
\label{equation_4}
\begin{aligned}
g_{a \gamma\gamma}\gtrsim 10^{-14}\,\mathrm{GeV^{-1}}  \times \left(\frac{\mathrm{SNR}}{Q}\right)^{1/2}   \times  \left(\frac{\mathrm{m}^2}{A}\right)^{1/2} \\ \times  \left(\frac{m_a}{\mathrm{\mu eV}}\right)
  \times \left(\frac{\mathrm{10\,s}}{t}\right)^{1/4} \times \left(\frac{T_\mathrm{{sys}}}{\mathrm{K}}\right)^{1/2}   \times \frac{\mathrm{10\,T}}{B_0} \\ \times  \underbrace{{\left(\frac{\delta\nu}{\mathrm{Hz}}\right)^{1/4}}}_{\mathrm{(i)}}   \times 
 \underbrace{f_{\mathrm{DM}}^{-1/2}}_{\mathrm{(ii)}} 
\,,
\end{aligned}
\end{equation}
where SNR is the signal to noise ratio, $t$ is the integration time and $T_{\mathrm{sys}}$ is the system temperature. In Eq. \ref{equation_4}, two relevant terms for this work are placed in the last line: \\(i) the $\delta \nu$ term. In practice, the accessible spectral resolution is bounded well by the signal linewidth caused by the velocity dispersion of the axions, well by the characteristics of the acquisition electronics when the signal linewidth is extremely narrow. Since SNR scales with $\delta\nu$, spectral dilution of the signal causes a lower SNR, detracting from the sensitivity to direct streams with respect to focused streams as inferred from Eq. \ref{equation_4}. However, in the case of data acquisition and processing in real experiments like DALI, the minimum bin width is $\delta \nu_\mathrm{b}=1/(2^N f_s)$. We typically set $2^N f_s\sim t$, with $t$ the data integration time, so $\delta \nu_\mathrm{b}\sim t^{-1}$ \cite{Cabrera:2023qkt}. On the other hand, the  1/f noise becomes negligible as long as $\delta G / G \ll 1/ \sqrt{\delta \nu \, t}$  \cite{1946RScI...17..268D}. Gain fluctuations are of the order of $\delta G/G\sim10^{-5}$ in high-electron-mobility transistor based radiometers, with a knee-frequency below 100 Hz \cite{2004SPIE.5470..402G}. We now consider two benchmarks from Table~\ref{tab:stream_table}: (A) a density over nominal of the focus region produced by Earth's gravity on a fine grained stream of $\rho_{\mathrm{str}}/\rho_0\sim10$ with $\sigma_{\mathrm{str}}\sim10^{-10}c$ and (B) a `direct' fine grained stream with $\rho_{\mathrm{str}}/\rho_0\sim10^{-2}$ and $\sigma_{\mathrm{str}}\sim10^{-17}c$. In the parameter space bounded by these two events, we have $10^{-35}\lesssim\delta\nu/\nu\lesssim10^{-21}$—with $\delta\nu/\nu\sim\sigma^2/2$—, and hence the radiometer is stabilized as long as $t<10^{18}$ s in the worst case, which is a period of time even longer than the age of the Universe. As a consequence, for an event lasting for $t \sim 10$ s, the resolution is limited by the DM event duration. In this picture, we have  $\delta \nu=\delta \nu_{\mathrm{b}}\sim 1/10$ Hz independently of the broadening of the signal caused by the velocity dispersion of the particles that make up the stream—obviously, this is in contrast to the usual case of the search for the Galactic axion in which the maximum integration time that makes the 1/f noise negligible for a linewidth of $\delta \nu / \nu \sim 10^{-6}$ is $t\ll 10$ ms per sub-spectrum that are stacked and averaged over periods of days to mitigate white noise. There could be alternative methods for tackling the data that we will investigate, perhaps allowing for increased resolution and, thus, sensitivity; \\(ii) the $f_{\mathrm{DM}}$ term that models the DM distribution. In the general case, the average energy density of an axion distribution reads
\begin{equation}
\label{equation_5}
\langle \rho_a \rangle=\frac{1}{V} \int \frac{d^3\mathrm{\textbf{p}}}{(2\pi)^3} \frac{m^2_a \lvert a(\mathrm{\textbf{p}}) \lvert ^2} {2} \, .
\end{equation}
\begin{figure}[h]
   \centering
 \vspace{2pt}\includegraphics [width=.5\textwidth]{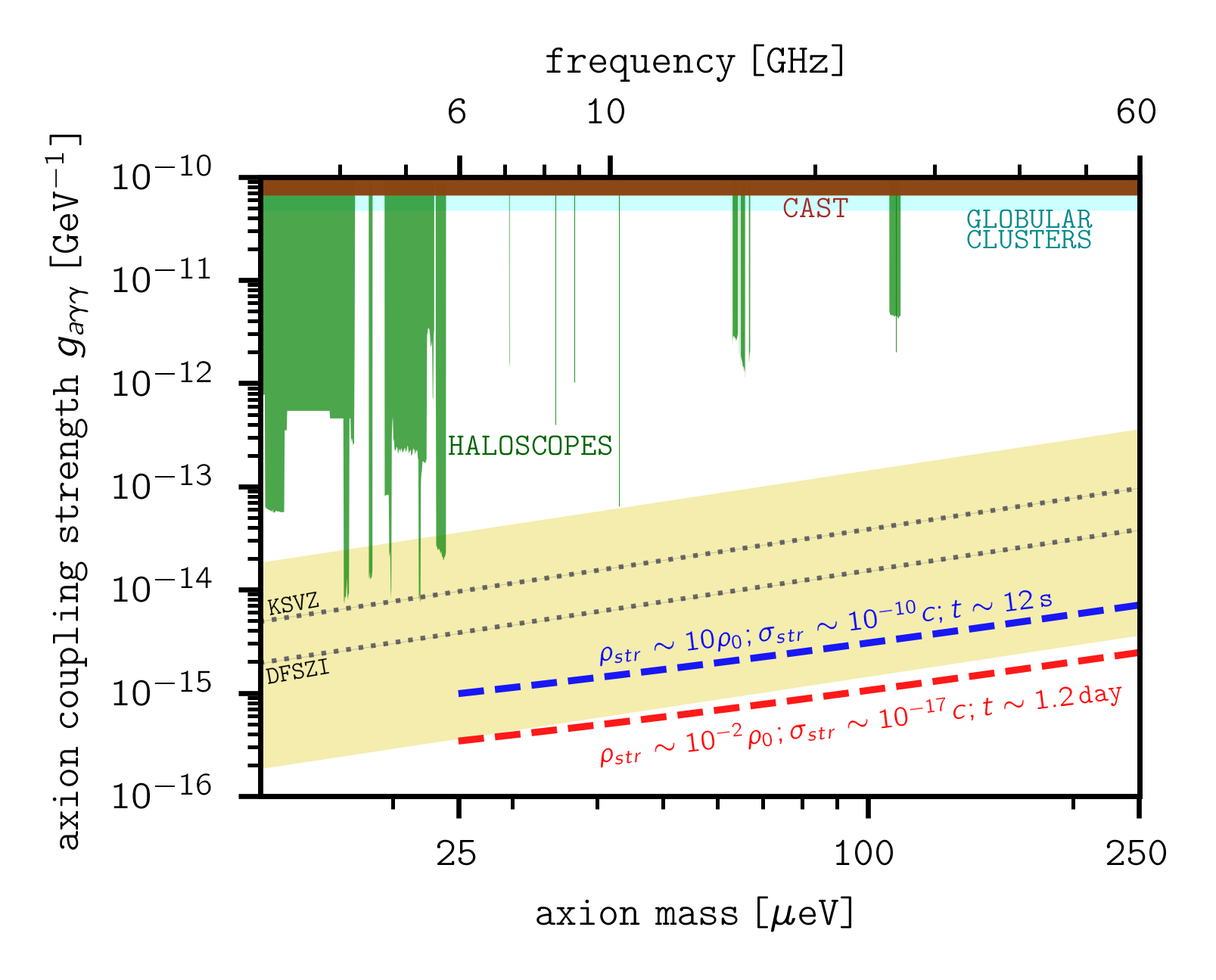}
\caption{Sensitivity projection for $\delta\nu\sim1/10$ Hz, bounded by DALI electronics, for gravitationally focused fine grained streams (case A, in blue) and not disturbed direct incident streams towards the DM axion detector system (B, in red). We adopt the benchmark parameters of DALI from \cite{DeMiguel:2023nmz}—$B_0\sim11.7$ T, $A\sim 3/2$ m$^{2}$, with N=50 layers. The system temperature is given by a physical temperature of about 1 K plus 3 times the quantum noise at each frequency; while we consider SNR $=5$, and $\rho_0\sim0.45$ GeV cm$^{-3}$. The instantaneous sweep bandwidth is $\pazocal{O}(100)$ MHz. The observation time $t\sim1.2$ day is constrained to a typical integration time per frequency step of DALI phase II. The plot shows the stacking of data over roughly 3.5 years of observations. Two representative QCD axion models are shown \cite{PhysRevLett.43.103, Shifman1980CanCE, DINE1981199, osti_7063072}. The yellow shaded region is compatible with the QCD axion.}
\label{fig}
\end{figure}

 In the case of a considerable inhomogeneity, each acquisition will observe a different energy density, thus resulting in a time modulation of the signal that may be introduced into $f_{\mathrm{DM}}$. However, if the DM energy density of the stream is approximately homogeneous over time scales of the order of $t$, the fluctuations vanish. In this case, $\langle \rho_{\mathrm{str}} \rangle \simeq f_{\mathrm{DM}}\rho_0$, which is the picture considered throughout this manuscript.

 Lastly, a negligible Doppler shift is caused by the rotation velocity of the Earth and the stream motion as $\nu'= \nu(c+v_{\oplus})/(c+v_{\mathrm{str}})$, where $v_{\oplus}$ is of the order of $10^{-6}c$. For directly investigated streams $v_{\mathrm{str}}\sim220\mathrm{ cos}\zeta$ km s$^{-1}$, while for GsF regions, $v_{\mathrm{str}}\sim20\mathrm{ cos}\zeta$ km s$^{-1}$~\cite{prdgf:2023}, with $\zeta$ as the incidence angle.
 A simulation is shown in Fig. \ref{fig}. For Fig. \ref{fig}, we employ Eq. \ref{equation_4} to study the discovery prospects with DALI for A and B events as defined previously in this section.
 

\section{Conclusions}\label{IV}
The introduced DM axions substructures in form of streams or clusters is a few decades old \cite{Tkachev:1991ka, PhysRevLett.71.3051, Hoffmann:2003zz, Sofue:2020mda, Prézeau_2015, Kryemadhi:2022vuk, 10.1111/j.1365-2966.2011.18224.x}. Although the initial purpose of DALI was to search for DM axions within the standard halo model, the same DALI set-up is simultaneously responsive also to transient DM axion events, e.g., streams or clusters. In the DALI projects, the same observational data are to be reduced by different pipelines with a different resolution. Even if the event rate of a collision, of the order of the few per cent, typically, is weighted by the frequency sweep step of DALI commensurating the probability of detection accordingly with the mass range compatible with axion DM—in other words, the instrument sequentially probes a few hundred megahertz of instantaneous bandwidth accommodated within several decades in frequency that are compatible with the DM axion but remain unexplored—with the present revised proposal, we address a new potential discovery channel aiming at the direct detection of DM axions covering the entire axion rest mass window as it is expected by 
some reference axion models. Thus, short-duration spectral features could eventually be explained.
In conclusion, DALI has a potential detection sensitivity by transient signatures, which so far have not been given the appropriate attention.

\section*{Acknowledgements}
We thank Juan F. Hernández-Cabrera, Elvio Hernández-Suárez, Enrique Jóven Álvarez, Haroldo Lorenzo-Hernández, Chiko Otani, and J. Alberto Rubiño-Martín, for discussions on DALI setup. The work of J.D.M. was supported by RIKEN's program for Special Postdoctoral Researchers (SPDR)—Project Code: 202101061013. The work of A.K. was supported by Messiah University's scholarship programs. We gratefully acknowledge financial support from the Severo Ochoa Program for Technological Projects and Major Surveys 2020-2023 under Grant No. CEX2019-000920-S; Recovery, Transformation and Resiliency Plan of Spanish Government under Grant No. C17.I02.CIENCIA.P5; Operational Program of the European Regional Development Fund (ERDF) under Grant No. EQC2019-006548-P; IAC Plan de Actuación 2022; Resident Astrophysicist Programme of the Instituto de Astrofísica de Canarias. 

\bibliography{apssamp}

\end{document}